\documentclass[prb,preprintnumbers,amsmath,amssymb,floatfix]{revtex4}
\usepackage{graphicx}
\usepackage{subfigure}
\usepackage{dcolumn}
\begin{document}

\title{A new one-dimensional variable frequency photonic crystal}
\author{ Xiang-Yao Wu$^{a}$\footnote{E-mail: wuxy2066@163.com}, Ji Ma$^{a}$,  Xiao-Jing
Liu$^{a}$, Hong Li$^{a}$ and Wan-Jin Chen$^{a}$ \footnote{E-mail:
chenwwjj@126.com}} \affiliation{$^{a}$ {\small Institute of
Physics, Jilin Normal University, Siping 136000, China}}

\begin{abstract}

In this paper, we have firstly proposed a new one-dimensional
variable frequency photonic crystal (VFPCs). We have calculated
the transmissivity and the electronic field distribution of VFPCs
and compare them with the conventional PCs, and obtained some new
results, which should be help to design a new type optical
devices, and the two-dimensional and three-dimensional VFPCs can
be studied further.

\vskip 10pt



\end{abstract}

\vskip 10pt \maketitle

 \vskip 8pt {\bf 1. Introduction} \vskip 8pt

Photonic crystal (PCs) were first introduced theoretically by
Yablonovitch [1]., and experimentally by John [2]. PCs,
constructed with periodic structure of artificial dielectrics or
metallic materials, have attracted many researchers in the past
two decades for their unique electromagnetic properties and
scientific and engineering applications [1-6]. These crystal
indicate a range of forbidden frequencies, called photonic band
gap, as a result of Bragg scattering of the electromagnetic waves
passing through such a periodical structure [7, 8]. As the
periodicity of the structure is broken by introducing a layer with
different optical properties, a localized defect mode will appear
inside the band gap. Enormous potential applications of PCs with
defect layers in different areas, such as light emitting diodes,
filters and fabrication of lasers have made such structures are
interesting research topic in this field.

Such materials are employed for the realization of diverse optical
devices, as for example distributed feedback laser [9, 10] and
optical switches [11, 12]. The addition of defects in the periodic
alternation, Or the realization of completely random sequences,
results in disordered photonic structures [13-15]. In the case of
one-dimensional disordered photonic structures, very interesting
physical phenomena Have been theoretically predicted or
experimentally observed.

In the paper, we have firstly proposed a new one-dimensional
variable frequency photonic crystal (VFPCs), which is made up of
variable frequency medium. The so-called variable frequency medium
is when light passes the medium the light frequency has been
changed, which is a new type medium, it changes the passed light
frequency. For the conventional medium, it changes the passed
light wavelength and unchanged light frequency. When the light
passes through the medium, the medium does not absorbs the light
energy, the light frequencies should be unchanged, the medium is
called conventional medium. When the light passes through the
medium, the medium absorbs or releases the light energy, the light
frequencies should be changed, the medium is called variable
frequency medium. We can make the photonic crystal with the
variable frequency medium. We have studied the transmissivity and
the electronic field distribution of VFPCs and compare them with
the conventional PCs. We obtained some new results: (1) When the
variable frequency function $f(n_b)<1$, the band gaps blue shift,
and the band gaps width decrease. (2) When the variable frequency
function $f(n_b)>1$, the band gaps red shift, and the band gaps
width increase, i.e., the variable frequency function is an
important factor of effect on transmissivity. (3) When the VFPCs
period numbers increase the band gaps width increase. In the
conventional PCs, the band gaps are unchanged when the period
numbers increase. (4) When the variable frequency function
$f(n_b)< 1$ the peak value of VFPCs electronic field distribution
decreased, the distribution curve left shift. (5) When the
variable frequency function $f(n_b)> 1$ the peak value of VFPCs
electronic field distribution increased, the distribution curve
right shift. We think these new results should be help to design a
new type optical devices.

 \vskip 8pt
 {\bf 2. Transfer matrix, transmissivity and electronic field distribution of PCs}
\vskip 8pt

For one-dimensional PCs, the calculations are performed using the
transfer matrix method [16], which is the most effective technique
to analyze the transmission properties of PCs. For the medium
layer $i$, the transfer matrices $M_i$ for $TE$ wave is given by
[16]:
\begin{eqnarray}
M_{i}=\left(%
\begin{array}{cc}
 \cos\delta_{i} & -i\sin\delta_{i}/\eta_{i} \\
 -i\eta_{i}sin\delta_{i}
 & \cos\delta_{i}\\
\end{array}%
\right),
\end{eqnarray}
where $\delta_{i}=\frac{\omega}{c} n_{i} d_i cos\theta_i$, $c$ is
speed of light in vacuum, $\theta_i$ is the ray angle inside the
layer $i$ with refractive index $n_i=\sqrt{\varepsilon_i \mu_i}$,
$\eta_i=\sqrt{\varepsilon_i/\mu_i} cos\theta_i$,
$cos\theta_i=\sqrt{1-(n^2_0sin^2\theta_0/n^2_i)}$, in which $n_0$
is the refractive index of the environment wherein the incidence
wave tends to enter the structure, and $\theta_0$ is the incident
angle.

The final transfer matrix $M$ for an $N$ period structure is given
by:
\begin{eqnarray}
\left(%
\begin{array}{c}
  E_{1} \\
  H_{1} \\
\end{array}%
\right)&=&M_{B}M_{A}M_{B}M_{A}\cdot\cdot\cdot M_{B}M_{A}\left(%
\begin{array}{c}
  E_{N+1} \\
  H_{N+1} \\
\end{array}%
\right)
\nonumber\\&=&M\left(%
\begin{array}{c}
  E_{N+1} \\
  H_{N+1} \\
\end{array}%
\right)=\left(%
\begin{array}{c c}
  A &  B \\
 C &  D \\
\end{array}%
\right)
 \left(%
\begin{array}{c}
  E_{N+1} \\
  H_{N+1} \\
\end{array}%
\right),
\end{eqnarray}
where
\begin{eqnarray}
M=\left(%
\begin{array}{c c}
  A &  B \\
 C &  D \\
\end{array}%
\right),
\end{eqnarray}
with the total transfer matrix $M$, we can obtain the transmission
coefficient $t$, and the transmissivity $T$, they are
\begin{eqnarray}
t=\frac{E_{N+1}}{E_{1}}=\frac{2\eta_{0}}{A\eta_{0}+B\eta_{0}\eta_{N+1}+C+D\eta_{N+1}},
\end{eqnarray}
\begin{eqnarray}
T=t\cdot t^{*}.
\end{eqnarray}
Where $\eta_{0}=\eta_{N+1}=\sqrt{\frac{\varepsilon_0}{\mu_0}}$.

The electronic field distribution at position $x$ is [16]
\begin{eqnarray}
\left(%
\begin{array}{c}
  E(x) \\
  H(x) \\
\end{array}%
\right)&&=M_{A}(a-x)M_{B}(b)(M_{A}(a)M_{B}(b))^{N-1}\left(%
\begin{array}{c}
  E_{N+1} \\
  H_{N+1} \\
\end{array}%
\right) \nonumber\\&&=\left(%
\begin{array}{cc}
 A'(x) & B'(x) \\
 C'(x) & D'(x)\\
\end{array}%
\right) \left(%
\begin{array}{c}
  E_{N+1} \\
  H_{N+1} \\
\end{array}%
\right),
\end{eqnarray}
with $E_{N+1}=E_{1}\cdot t$ and
$H_{N+1}=\sqrt{\varepsilon_0/\mu_0} \cdot E_{N+1}$, we have
\begin{eqnarray}
E(x)=(A'(x)+B'(x)\sqrt{\varepsilon_0/\mu_0})E_{1}\cdot t,
\end{eqnarray}
and
\begin{eqnarray}
|\frac{E(x)}{E_{1}}|^2= |A'(x)+B'(x)\sqrt{\varepsilon_0/\mu_0}|^2
\cdot|t|^2.
\end{eqnarray}
\vskip 8pt {\bf 3. Transform matrix and transmissivity of
VFPCs}\vskip 8pt

In the following, we have firstly proposed a new one-dimensional
photonic crystal of variable frequency medium (VFPCs). The
so-called variable frequency medium is when light passes the
medium the light frequency has been changed, it is
\begin{eqnarray}
\omega=f(n)\omega_0,
\end{eqnarray}
where $\omega_0$ is the incident light frequency, $\omega$ is the
frequency of light in the medium and $n$ is the refractive index
of variable frequency medium, and $f(n)$ is called variable
frequency function. The structure of one-dimensional VFPCs is
$(AB)^N$, where the medium $B$ is the variable frequency medium,
the medium $A$ is not the variable frequency medium
(conventional medium), and $N$ is the period numbers\\

For the VFPCs, its transfer matrices are similar to the
conventional transfer matrices (1), they are
\begin{eqnarray}
M_{A_i}=\left(%
\begin{array}{cc}
 \cos\delta_{a_i} & -i\sin\delta_{a_i}/\eta_{a} \\
 -i\eta_{a}sin\delta_{a_i}
 & \cos\delta_{a_i}\\
\end{array}%
\right)(i=1, 2, \cdot\cdot\cdot, N),
\end{eqnarray}
\begin{eqnarray}
M_{B_i}=\left(%
\begin{array}{cc}
 \cos\delta_{b_i} & -i\sin\delta_{b_i}/\eta_{b} \\
 -i\eta_{b}sin\delta_{b_i}
 & \cos\delta_{b_i}\\
\end{array}%
\right)(i=1, 2, \cdot\cdot\cdot, N),
\end{eqnarray}
but the phases should be modified as:
\begin{eqnarray}
\delta_{a_1}=\frac{\omega_0}{c}\cdot n_{a}\cdot a,
\end{eqnarray}
\begin{eqnarray}
\delta_{b_1}=\frac{f(n_b)\cdot \omega_0}{c}\cdot n_{b}\cdot b,
\end{eqnarray}
\begin{eqnarray}
\delta_{a_2}=\frac{f(n_b)\cdot \omega_0}{c}\cdot n_{a}\cdot a,
\end{eqnarray}
\begin{eqnarray}
\delta_{b_2}= \frac{f^2(n_b)\cdot \omega_0}{c}\cdot n_{b}\cdot b,
\end{eqnarray}
and the i-th period phase is
\begin{eqnarray}
\delta_{a_i}=\frac{f^{i-1}(n_b)\cdot \omega_0}{c}\cdot n_{a}\cdot
a,
\end{eqnarray}
\begin{eqnarray}
\delta_{b_i}= \frac{f^{i}(n_b)\cdot \omega_0}{c}\cdot n_{b}\cdot
b,
\end{eqnarray}
We consider the structure of VFPCs is $(AB)^{12}$, and the
variable frequency function is $f(n_b)$. When the $f(n_b)>1$, the
light frequency should be increased in medium $B$, when the
$f(n_b)<1$, the light frequency should be decreased in medium $B$.
When $f(n_b)=1$ it is called conventional PCs, and $f(n_b)\neq 1$
it is called variable frequency PCs (VFPCs).

 \vskip 8pt
 {\bf 4. Numerical result}
 \vskip 8pt
In this section, we report our numerical results of VFPCs. The
VFPCs main parameters are: The medium $A$ refractive indexes
$n_a=1.40$, and thickness $a=320nm$, the medium $B$ refractive
indexes $n_b=2.58$, thickness $b=138nm$, The structure is
$AB^{12}$. In FIG. 1, we study the effect of the variable
frequency function $f(n_b)$ on VFPCs transmissivity, the variable
frequency function $f(n_b)$ in the FIG. 1 (a), (b) and (c) are
$0.98$, $1$ and $1.02$, respectively. The FIG. 1 (b) is the
transmissivity of conventional PCs because of $f(n_b)=1$, and FIG.
1 (a) and (c) are the transmissivity of VFPCs because of $f\neq
1$. Comparing the transmissivity of VFPCs with conventional PCs,
we can obtain the new results: (1) When the variable frequency
function $f(n_b)<1$ (FIG. 1 (a)), the band gaps blue shift, and
the band gaps width decrease. (2) When the variable frequency
function $f(n_b)>1$ (FIG. 1 (c)), the band gaps red shift, and the
band gaps width increase. From the results (1) and (2), we find
the variable frequency function is an important factor of effect
on transmissivity. In FIG. 2, we study the effect of the variable
frequency medium thickness on the VFPCs transmissivity, wherein
variable frequency function $f(n_b)=1.02$. The FIG. 2 (a), (b) and
(c) thickness $b$ are $98nm$, $138nm$ and $178nm$, respectively.
From FIG. 2 (a), (b) and (c), we can find when the variable
frequency medium thickness b increase the band gaps red shift, and
the band gaps width increase. In FIG. 3, we study the effect of
the variable frequency medium refractive index on the VFPCs
transmissivity, wherein variable frequency function $f(n_b)=1.02$.
The FIG. 3 (a), (b) and (c) thickness $n_b$ are $2.18$, $2.58$ and
$2.98$, respectively. From FIG. 3 (a), (b) and (c), we can find
when the variable frequency medium refractive index $n_b$ increase
the band gaps red shift, and the band gaps width increase. In FIG.
4, we study the effect of the period number $N$ on the VFPCs
transmissivity, wherein variable frequency function $f(n_b)=1.02$.
The FIG. 4 (a), (b) and (c) period number $N$ are $8$, $10$ and
$12$, respectively. From FIG. 4 (a), (b) and (c), we can find when
the VFPCs period numbers increase the band gaps width increase. In
the conventional PCs, the band gaps are unchanged when the period
numbers increase. In FIG. 5, we calculate the electronic field
distribution of the VFPCs, and consider the influence of variable
frequency function $f(n_b)$ on VFPCs electronic field
distribution. The variable frequency function $f(n_b)=0.97$, $1$
and $1.03$ are according to the dash dot line, solid line and dot
line of electronic field distribution, respectively. comparing
with the conventional PCs ($f(n_b)=1$), we can obtain new results:
(1) When the variable frequency function $f(n_b)< 1$ the peak
value of VFPCs electronic field distribution decreased, the
distribution curve left shift. (2) When the variable frequency
function $f(n_b)> 1$ the peak value of VFPCs electronic field
distribution increased, the distribution curve right shift.

{\bf 5. Conclusion}
 \vskip 8pt
In summary, we have studied the transmissivity and the electronic
field distribution of one-dimensional VFPCs and compare them with
the conventional PCs. We obtained some new results: (1) When the
variable frequency function $f(n_b)<1$, the band gaps blue shift,
and the band gaps width decrease. (2) When the variable frequency
function $f(n_b)>1$, the band gaps red shift, and the band gaps
width increase, i.e., the variable frequency function is an
important factor of effect on transmissivity. (3) When the VFPCs
period numbers increase the band gaps width increase. In the
conventional PCs, the band gaps are unchanged when the period
numbers increase. (4) When the variable frequency function
$f(n_b)< 1$ the peak value of VFPCs electronic field distribution
decreased, the distribution curve left shift. (5) When the
variable frequency function $f(n_b)> 1$ the peak value of VFPCs
electronic field distribution increased, the distribution curve
right shift. We think these new results should be help to design a
new type optical devices and we can further study the
two-dimensional and three-dimensional VFPCs.

\newpage
\begin{figure}[tbp]
\includegraphics[width=9 cm]{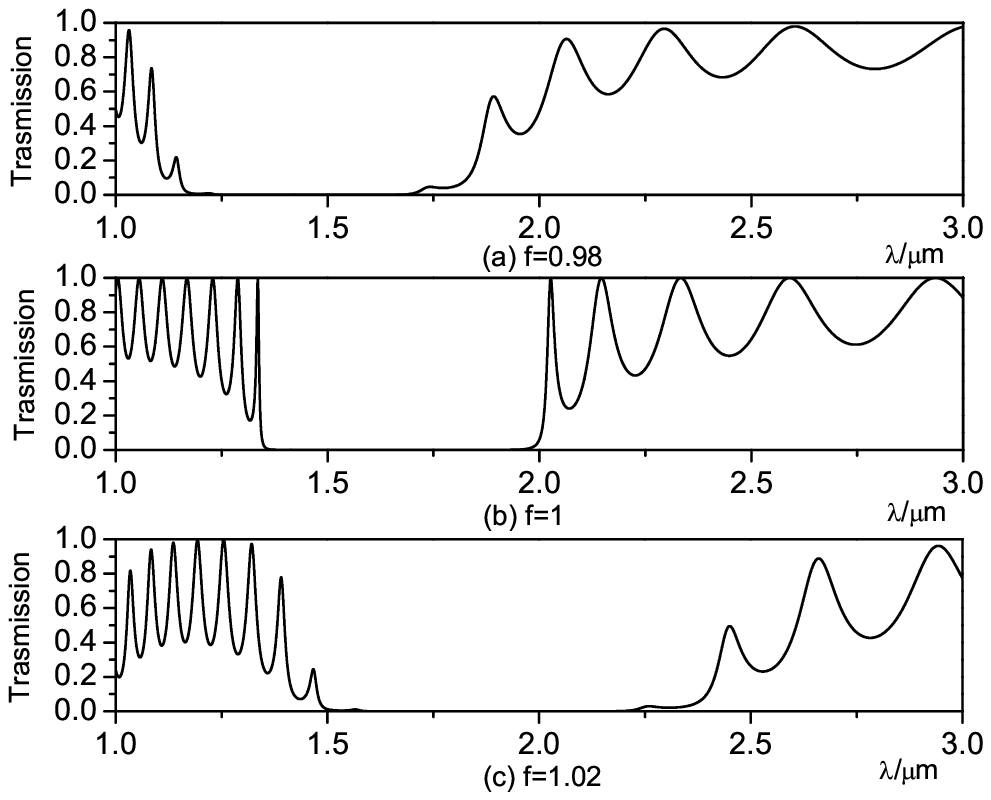}
\caption{Comparing the transmissivity of VFPCs with conventional
PCs. (a) VFPCs $f(n_b)=0.98$, (b) conventional PCs $f(n_b)=1$, (c)
VFPCs $f(n_b)=1.02$.}
\end{figure}

\begin{figure}[tbp]
\includegraphics[width=9 cm]{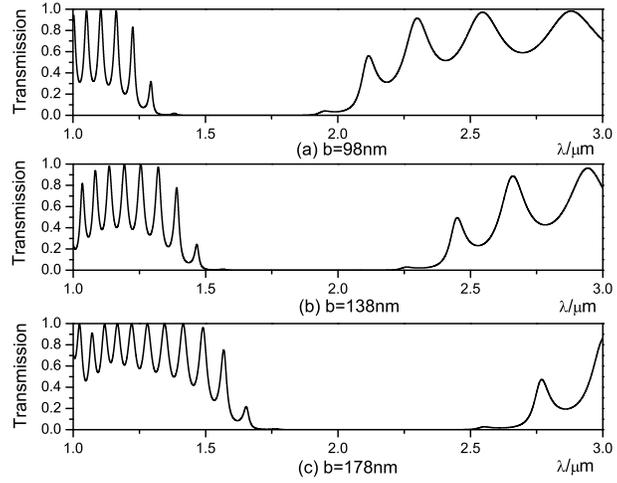}
\caption{The effect of the variable frequency medium thickness $b$
on the VFPCs transmissivity. (a) $b=98nm$, (b) $b=138nm$, (c)
$b=178nm$.}
\end{figure}

\begin{figure}[tbp]
\includegraphics[width=9 cm]{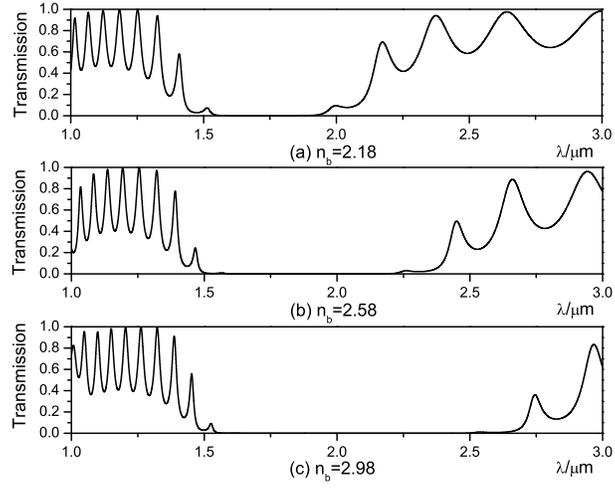}
\caption{The effect of the variable frequency medium refractive
index $n_b$ on the VFPCs transmissivity. (a) $n_b=2.18$, (b)
$n_b=2.58$, (c) $n_b=2.98$.}
\end{figure}

\begin{figure}[tbp]
\includegraphics[width=9 cm]{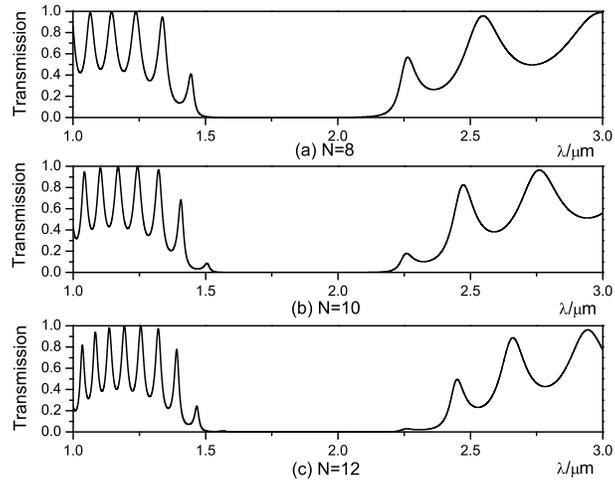}
\caption{The effect of the period number $N$ on the VFPCs
transmissivity. (a) $N=8$, (b) $N=10$, (c) $N=12$.}
\end{figure}

\begin{figure}[tbp]
\includegraphics[width=16 cm]{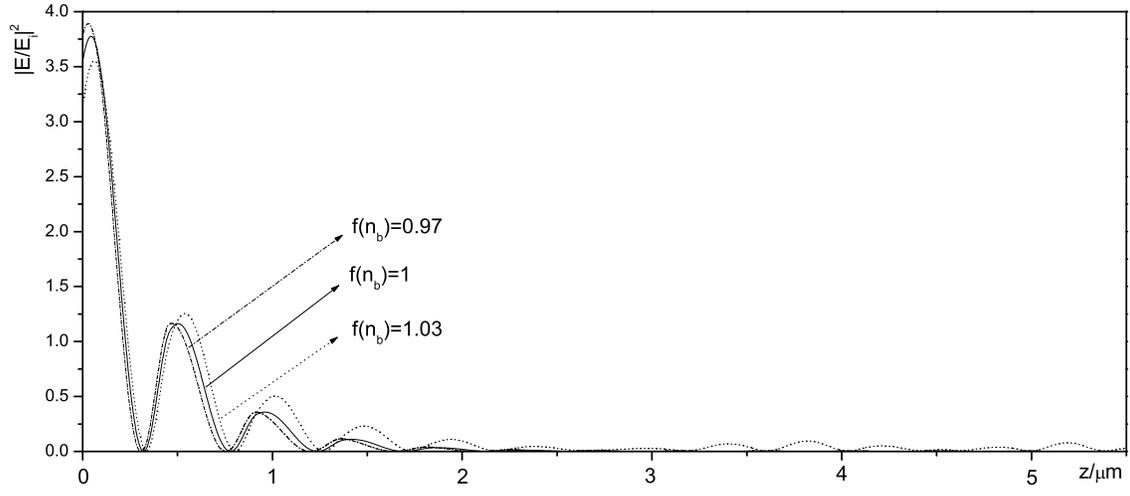}
\caption{The effect of the variable frequency function $f(n_b)$ on
the VFPCs electronic field distribution $|E/E_i|^2$. (a)
$f(n_b)=0.97$ dash dot line, (b) $f(n_b)=1$ solid line, (c)
$f(n_b)=1.03$ dot line.}
\end{figure}

\end{document}